\documentclass[twocolumn,showpacs,preprintnumbers,amsmath,amssymb,pra]{revtex4}
\pdfoutput=1

\usepackage{graphicx}
\usepackage{dcolumn}
\usepackage{bm}
\usepackage[usenames]{xcolor}

\begin{document}

\newcommand{\tr}{{\rm Tr\thinspace}}
\newcommand{\bra}[1]{\left\langle{#1}\right\vert}
\newcommand{\ket}[1]{\left\vert{#1}\right\rangle}
\newcommand{\ketil}[1]{\vert{#1}\rangle}
\newcommand{\ketbra}[2]{\left| #1 \right\rangle\!\!\!\,\left\langle #2 \right|}
\newcommand{\abs}[1]{\left\vert #1 \right\vert}
\newcommand{\expect}[1]{\left\langle{#1}\right\rangle}
\newcommand{\timeorder} {\underset{\leftarrow}{\mathcal{T}}}
\newcommand{\ident}{{\mathbb1}}
\newcommand{\order}[1]{\mathcal{O}\!\left( #1 \right)}
\newcommand{\diag}[1]{\mathrm{diag}\{#1\}}
\newcommand{\trans}[1]{#1^\mathsf{T}}
\newcommand{\T}{\mathsf{T}}
\newcommand{\N}{\mathcal{N}}

\newcommand{\Lket}[1]{\left|#1\middle\rangle\!\right\rangle}
\newcommand{\Lbra}[1]{\left\langle\!\middle\langle#1\right|}
\newcommand{\Lbraket}[2]{\left\langle\!\middle\langle #1 \middle| #2 \middle\rangle\!\right\rangle}
\newcommand{\Lketbra}[2]{\left| #1 \middle\rangle\!\middle\rangle \middle\langle\!\middle\langle #2 \right|}

\title{Collective Effects in Linear Spectroscopy of Dipole-Coupled Molecular Arrays}
     
\author{A. A. Kocherzhenko$^1$, J. Dawlaty$^2$, B. P. Abolins$^1$, F. Herrera$^{3,4}$, D. B. Abraham$^5$, K. B. Whaley$^1$}
\affiliation{$^1$Department of Chemistry, University of California, Berkeley, CA 94720\\
$^2$Department of Chemistry, University of Southern California, Los Angeles, CA 90089\\
$^3$Department of Chemistry and Chemical Biology, Harvard University, Cambridge, MA 02138\\
$^4$Department of Chemistry, Purdue University, West Lafayette, IN 47907\\
$^5$Department of Theoretical Physics, University of Oxford, UK}
\date{\today}

\begin{abstract}
We present a consistent analysis of linear spectroscopy for arrays of nearest neighbor dipole-coupled two-level molecules that reveals distinct signatures of weak and strong coupling regimes separated for infinite size arrays by a quantum critical point. In the weak coupling regime, the ground state of the molecular array is disordered, but in the strong coupling regime it has (anti)ferroelectric ordering.  We show that multiple molecular excitations (odd/even in weak/strong coupling regime) can be accessed directly from the ground state. We analyze the scaling of absorption and emission with system size and find that the oscillator strengths show enhanced superradiant behavior in both ordered and disordered phases. As the coupling increases, the single excitation oscillator strength rapidly exceeds the well known Heitler-London value. In the strong coupling regime we show the existence of a unique spectral transition with excitation energy that can be tuned by varying the system size and that asymptotically approaches zero for large systems. The oscillator strength for this transition scales quadratically with system size, showing an anomalous one-photon superradiance. For systems of infinite size, we find a novel, singular spectroscopic signature of the quantum phase transition between disordered and ordered ground states. We outline how arrays of ultra cold dipolar molecules trapped in an optical lattice can be used to access the strong coupling regime and observe the anomalous superradiant effects associated with this regime.
\end{abstract}

\maketitle

\section{Introduction}

Quantum correlations between light-absorbing pigments can enhance light-matter interaction, resulting in superradiance, i.e., in extraordinarily fast spontaneous emission~\cite{spano1989superradiance,fidder1991optical}. In most materials where superradiance is observed, the interaction energy between chromophores is smaller than the single-chromophore excitation energy. However, systems with interactions that significantly exceed those typical in organic chromophore aggregates are common in today's quantum technology. Arrays with intersite interaction strengths that are comparable to or greater than the single-site excitation energy can be emulated using ion traps~\cite{Kim2011,Islam03052013,Senko25072014, islam2011onset}, superconducting circuits~\cite{Johnson2011}, and optical traps for neutral atoms~\cite{Simon2011} or molecules~\cite{Micheli2006,gorshkov2011quantum,Gorshkov2011,herrera2014entanglement}. Systems with stronger interactions between sites can form a correlated quantum phase, separated for infinite systems from the weak interaction regime by a quantum critical point. We propose arrays of ultra cold dipolar molecules trapped in an optical lattice as a candidate system where the strong coupling regime can be emulated. We show that this strong coupling regime is characterized by qualitatively new spectroscopic properties, including an anomalous one-photon superradiance that scales quadratically with system size. Development of materials that possess such properties opens new prospects for the efficient capture and sensing of light.

Consider a one-dimensional array of $M$ two-level chromophores with excitation energy $\varepsilon$ that are coupled by dipole-dipole interactions $b$ between nearest neighbors. This system is described by the Hamiltonian of Krugler, Montgomery and McConnell (KMM)~\cite{krugler1964collective}:
\begin{equation}
H= \sum_{m=1}^M \left[ \varepsilon P_m^\dagger P_m + b ( P_m^\dagger + P_m ) ( P_{m+1}^\dagger + P_{m+1}) \right] .
\label{eq:H_KMM}
\end{equation}
Here $P_m^\dagger$ creates and $P_m$ annihilates an excitation at site $m$ (we work with unit lattice spacing). Each of these operators is a product of a pair of electron creation and annihilation operators in the molecular basis. Since charge transfer is not allowed, pairs of the operators $P_m^\dagger$, $P_m$ commute off site; on site we have $P_m^\dagger P_m + P_m P_m^\dagger = 1$.

KMM found significant co-operative effects in the ground state of Eq.~(\ref{eq:H_KMM}). As $B = 2|b| / \varepsilon$ is increased through unity, a transition from a non-degenerate to a two-fold degenerate ground state is seen for systems of infinite size. KMM conjectured that this reflects a transition from a ground state with no electronic polarization to one showing electronic (anti)ferroelectric polarization, resulting from long-range (anti)ferroelectric ordering of the transition dipole moments. This corresponds to the transition from the paramagnetic to the (anti)ferromagnetic phase for the equivalent quantum Ising Hamiltonian:
\begin{equation}
H_{\mathrm{spin}} = \varepsilon \sum_{m=1}^M (1+\sigma_m^z)/2 + b \sum_{m=1}^M \sigma_m^x \sigma_{m+1}^x,
\label{eq:ising}
\end{equation}
that is obtained by treating $P_m$ as a spin lowering operator for a chain of spin-1/2 entities, i.e., $P_m^\dagger + P_m = \sigma_m^x$. Quantum fluctuations do not destroy the ordering in the Hamiltonian ground state at $B > 1$: this was first confirmed for $XY$ models with $Z$ magnetic fields in the seminal work of McCoy~\cite{McCoy1968}. For $M \rightarrow \infty$, the ordered regime at strong couplings is separated from the disordered regime at weak couplings by a quantum critical point at $B=1$~\cite{pfeuty1970one}. From now on we will restrict our attention to the case $b<0$~\cite{note_othercases}. 

For crystals and aggregates of molecular chromophores, the dipole-dipole coupling $\left| b \right|$ is typically smaller than the excitation energy $\varepsilon$ by an order of magnitude or more~\cite{bakalis1997optical}, so that $B \ll 1$. Studies of excitonic energy transfer and spectroscopy of such systems in this very weak coupling limit generally invoke the Heitler-London (HL) approximation to Eq.~(\ref{eq:H_KMM}). This approximation ignores the double excitation, $P^\dagger_m P^\dagger_n$, and double de-excitation, $P_m P_n$, terms in the Hamiltonian, despite the matrix elements for these terms in Eq.~(\ref{eq:H_KMM}) being equal to those for the hopping terms, $P^\dagger_m P_n$ and $P_m P^\dagger_n$. The HL approximation adequately reproduces observed optical spectra for $B \ll 1$~\cite{bakalis1997optical}, but ignores collective effects in the ground state, and is thus intrinsically inconsistent. This can be illustrated by considering the KMM Hamiltonian for $M = 2$. Its matrix representation in the basis of isolated chromophore states is block-diagonal with two $2 \times 2$ blocks: one involving the single excitation states, with eigenvalues $\varepsilon \pm b$, the other involving states with zero and two excitations, with eigenvalues $\varepsilon \pm \left[ \varepsilon^2 + b^2 \right]^{1/2}$. On the other hand, in the HL approximation to the Hamiltonian, the zero-excitation and two-excitation states are uncoupled, yielding eigenvalues $0$ and $2\varepsilon$, which are identical to those of two non-interacting chromophores and are clearly incorrect.

Several authors have analyzed the differences from HL spectra that arise when double excitation and deexcitation terms are included in Eq.~(\ref{eq:H_KMM}), for the limit of very weak coupling, $B \ll 1$~\cite{spano1991fermion,bakalis1997optical,agranovich2000frenkel}. Such non-HL terms have also been shown to play a role in the spectroscopy of small clusters of polar molecules, which are described by considerably more complex Hamiltonians than Eq.~(\ref{eq:H_KMM})~\cite{painelli2003multielectron, terenziani2003supramolecular}. Prior studies of spectra deriving from Eq.~(\ref{eq:H_KMM}) did not address the interesting strong coupling regime of (anti)ferroelectric ground state polarization.  Furthermore, they focused either on numerical calculations for finite systems with fixed boundary conditions~\cite{spano1991fermion,bakalis1997optical} or on perturbative analysis for periodic boundary conditions~\cite{agranovich2000frenkel}. Consequently, no prior work has revealed any signature of a quantum phase transition, for which exact analysis of the spectroscopic response of an infinite chain is required. 

Modern quantum technology offers possibilities for analog simulation of the transverse Ising Hamiltonian, Eq.~(\ref{eq:H_KMM}), for a wide range of $B$ values~\cite{Kim2011, Johnson2011, Simon2011, Micheli2006}. This motivates interest in the spectroscopic properties of both the paraelectric ($B < 1$) and the (anti)ferroelectric ($B > 1$) phases that can be realized in such systems.

While KMM predicted the energetics and the structure of eigenstates for systems described by Eq.~(\ref{eq:H_KMM})~\cite{krugler1964collective}, they did not determine the optical spectra that result from this Hamiltonian. The present study addresses this key issue by taking advantage of new, relevant techniques in order to develop an exact solution for the linear absorption spectrum at arbitrary $B$ values for both infinite and finite size systems. Since the eigenstates predicted by the HL approximation are incorrect for all but extremely small couplings, the spectroscopic analysis of Eq.~(\ref{eq:H_KMM}) reveals spectroscopic behavior that cannot be described within the HL approximation, both for strong coupling ($B > 1$) and weak  coupling ($B < 1$) regimes. For $0 < B < 1$, we find that optical transitions between the ground state and states with any odd number of excitations may be observed, while for $B > 1$ optical transitions between the ground state and states with even excitation numbers are permitted. We analyze the scaling of absorption and emission with system size and find that the oscillator strengths show enhanced superradiant behavior in both ordered and disordered phases. Of particular note is the finding of a quadratic scaling of absorption with system size in the strong coupling regime, corresponding to enhanced emission that exceeds the well-known linear scaling of one-photon superradiance seen for both non-interacting systems~\cite{dicke1954coherence,scully2009super} and molecular aggregates in the Heitler-London limit~\cite{agranovich1999quantum,knoester2007modeling}.This change from linear to quadratic scaling of one-photon absorption and emission constitutes a one-photon analog of the anomalous size scaling of superradiance, termed hyperradiance, that is seen in phase-locked soliton oscillators~\cite{gronbech1993hyperradiance}. Our analysis of infinite systems, $M \longrightarrow \infty$, reveals a novel, singular spectroscopic signature of the quantum phase transition between disordered and ordered ground states. Finally, for $B > 1$, we show the existance of a transition with excitation energy that can be tuned by varying the system size $M$ and that asymptotically approaches zero as $M \longrightarrow \infty$. 

In Sections~\ref{sec:reviewofeigenstatecalculation} -- \ref{sec:linearspectroscopy} we present the general spectroscopic analysis of Eq.~(\ref{eq:H_KMM}). In Section \ref{sec:realization}, we outline an implementation of the Hamiltonian given by Eq.~(\ref{eq:H_KMM}) that uses arrays of ultra cold dipolar molecules trapped in an optical lattice. This implementation permits access to the strong coupling regime, $B > 1$, and the observation of hyperradiant effects associated with this regime. Section~\ref{sec:summary} provides a summary and conclusions.
 
\section{Review of eigenstate calculation} 
\label{sec:reviewofeigenstatecalculation}

Since the eigenstates of Eq.~(\ref{eq:H_KMM}) are critical for our spectroscopic analysis, we first summarize the key features of the analytic diagonalization of this Hamiltonian that was carried out by KMM~\cite{krugler1964collective}.
The first step is a Jordan-Wigner transformation of $\{ P_m \}$:
\begin{equation}
f_1 = P_1,~f_m = Q_{m-1} P_m,~Q_m = \prod_{j=1}^m \left( 1 - 2 P_j^\dagger P_j \right),
\end{equation}
for $ 2 \le m \le M$.

The operators $\left\lbrace f_m, f_m^\dagger \right\rbrace$ have lattice-fermionic anti-commutation relations, so Eq.~(\ref{eq:H_KMM}) becomes a quadratic form in fermions, except for the boundary term, which gives the expected form in fermions, but multiplied by $-Q_M$. Since $\left[ H, Q_M \right] = 0$, the Hamiltonian can be decomposed into two quadratic forms, $H_{+}$ and $H_{-}$, by projection onto orthogonal subspaces: $H = Q_{+}H_{+} + Q_{-}H_{-}$, where $Q_{\pm} = \left( 1 \pm Q_M \right) /2$. $H_{+}$ and $H_{-}$ are then diagonalized separately, after applying a discrete Fourier transformation: 
\begin{equation}
\label{eq:Fourier}
F^\dagger \left( k \right) = M^{-1/2} \sum_{m = 1}^M e^{i k m} f_m^\dagger,
\end{equation}
where for $H_{\pm}:~ e^{ikM} = \mp 1$, $0 \le k < 2\pi$. The allowed values of the wavenumber $k$  are denoted $\alpha = 2\pi \left( m - 1 \right)/M$ for $H_{-}$ and $\beta = \pi \left( 2 m - 1 \right)/M$ for $H_{+}$.

The Bogoliubov-Valatin transformation
\begin{equation}\label{eq:BogoliubovValatin}
G^\dagger \left( k \right) = \cos \theta(k) F^\dagger(k) - i \sin \theta(k) F(-k)
\end{equation}
diagonalizes $H_{+}$($H_{-}$) for $k = \beta(\alpha)$, yielding
\begin{equation}\label{eq:diagHam}
H_{\pm} = E_{\pm} + \sum_{\substack {k:~\exp\left(i k M\right) = \mp 1, \\ 0 \leq k < 2\pi}} E \left( k \right) G^\dagger \left( k \right) G \left( k \right),
\end{equation}
where
\begin{equation}\label{eq:theta}
\tan{\theta(k)} = \left( 2 b \sin k \right)^{-1} \left[ E(k) - E_0(k) \right],
\end{equation}
\begin{equation}\label{eq:excenergy}
E(k) = \left[E_0(k)^2 + 4b^2 \sin^2 k \right]^{1/2},
\end{equation}
and $E_0(k) = \varepsilon+2b\cos k$ is the HL dispersion relation.

The ground states $\left| \Phi_{\pm} \right\rangle$ of $H_{\pm}$ resemble the Bardeen-Cooper-Schrieffer (BCS) ground state and have energies~\cite{krugler1964collective}
\begin{equation}\label{eq:gsenergy}
E_{\pm} = -\frac{1}{2} \sum_{\substack {k:~\exp\left(i k M\right) = \mp 1, \\ 0 \leq k < 2\pi}} \left[ E \left( k \right) - E_0 \left( k \right) \right].
\end{equation}
All eigenstates of Eq.~(\ref{eq:H_KMM}) are then given by the eigenstates of $H_{+}$ produced by applying an even number of $G^\dagger \left( \beta \right)$ operators to $\left| \Phi_{+} \right\rangle$, together with the eigenstates of $H_{-}$ produced by applying an odd ($B<1$) or even ($B>1$) number of $G^\dagger \left( \alpha \right)$ operators to $\left| \Phi_{-} \right\rangle$ (see Appendix~\ref{sec:translationalsymmetry}). The ground state of $H$ is always $\left|\Phi_{+}\right\rangle$.

\section{Exact calculation of dipole matrix elements for finite arrays}

Consider the interaction of the chromophore arrays studied here with light, in the electric dipole approximation. Since the dipole excitation operator $\sigma_m^x = P_m^\dagger + P_m$ anti-commutes with parity, some immediate predictions about the linear spectra can be made.

First, the ground state is coupled to \textit{all} states of opposite parity. This is in contrast with the HL approximation, for which the only allowed transitions from the ground state are to single-excitation states~\cite{spano1991fermion,bakalis1997optical}.

Second, from Eq.~(\ref{eq:gsenergy}) it follows that for $B > 1$ the energy of the lowest excitation $\left| \Phi_{+} \right\rangle \longrightarrow \left| \Phi_{-} \right\rangle$ is
\begin{equation}\label{eq:lowestexcen}
E_{-} - E_{+} = \frac{1}{2} \left[ \sum_{0 \leq \beta < 2\pi} E \left( \beta \right) - \sum_{0 \leq \alpha < 2\pi} E \left( \alpha \right) \right],
\end{equation}
since $E_0 \left( k \right)$ is an even function and, consequently,
\begin{equation}
\sum_{0 \leq \alpha < 2\pi} E_0 \left( \alpha \right) - \sum_{0 \leq \beta < 2\pi} E_0 \left( \beta \right) = 0.
\end{equation}
Eq.~(\ref{eq:lowestexcen}) can be evaluated as
\begin{multline}\label{eq:vactovacexcen}
E_{-} - E_{+} = \frac{2 M \varepsilon \left( 2 B \right)^{1/2}}{\pi} \cdot \\ \cdot \int_{v_0}^\infty \mathrm{d}v \frac{e^{-M v}}{1 - e^{- 2 M v} } \left( \cosh v - \cosh v_0 \right)^{1/2},
\end{multline}
where $\cosh v_0 = \left( B + B^{-1} \right) / 2$ (see Appendix~\ref{sec:lowestexceninstrongcoup}).
From Eq.~(\ref{eq:vactovacexcen}) it follows that as $M \longrightarrow \infty$, the lowest excitation energy $E_{-} - E_{+} \longrightarrow 0$. Provided the matrix element for this transition is non-zero, photon absorption at arbitrarily low frequencies is expected for an array of strongly coupled chromophores.

Third, it should be noted that for one-dimensional systems, quantum fluctuations do not necessarily destroy long range order, but a system dimensionality of two or greater is needed to stabilize ordered states against thermal fluctuations. Writing $\ket{\Phi_{\pm}}=\left( \ket{+} \pm \ket{-} \right)/\sqrt{2}$, where $P_M\ket{+}=\ket{-}$ and $\bra{+} \sigma_1^x \ket{+} \rightarrow \left( 1 - 1/B^2 \right)^{1/8}$ as $M\rightarrow \infty$, we see that excitations are formed by introducing domains of reversed polarization that result from applying pairs of local ``flip" operators. Excitations are thus generated in pairs in the strong coupling regime: the fact that only even numbers are allowed here is a topological constraint imposed by the periodic boundary conditions (see Appendix~\ref{sec:translationalsymmetry}). 

We now describe the calculation of transition matrix elements in the strong coupling regime, $B > 1$. Without loss of generality, we restrict our attention to $\bra{\Phi_{-}} G\left(\alpha_{2n}\right) \dots G\left(\alpha_1\right) \sigma_1^x \left| \Phi_{+} \right\rangle$, since all allowed excitations may be generated from $\sigma_1^x$ by making use of the translational symmetry of Eq.~(\ref{eq:H_KMM}): $TP_mT^\dagger = P_{m-1}, 2 \le m \le M$ with $TP_1T^\dagger = P_M$.
In the strong coupling regime, the Hamiltonian ground state, $\ket{\Phi_{+}}$, has even parity and the ground state of $H_{-}$, $\ket{\Phi_{-}}$, has odd parity. 
Consequently, the allowed optical transitions from $\ket{\Phi_{+}}$ are to $\ket{\Phi_{-}}$ and to states with an even number of excitations generated from the latter state, with corresponding transition dipole moments $ \bra{\Phi_{-}} G\left(\alpha_{2n}\right) \dots G\left(\alpha_1\right) \sigma_1^x \ket{ \Phi_{+}}$. Using an extension of Wick's theorem~\cite{Abraham2012}, these matrix elements can be shown to satisfy
\begin{widetext}
\begin{equation}
\begin{split}
\sum_{\alpha_1} 
(\alpha_1, \beta)_1 \bra{\Phi_{-}} G\left(\alpha_{2n}\right) \dots G\left(\alpha_1\right) \sigma_1^x \ket{ \Phi_{+}} = \sum_{j=2}^{2n} (-1)^{j-1} (-\alpha_j, \beta)_2 \Delta_{1j} \bra{\Phi_{-}} G\left(\alpha_{2n}\right) \dots G\left(\alpha_1\right) \sigma_1^x \ket{ \Phi_{+}}, 
\label{eq:strongcouplingWick}
\end{split}
\end{equation}
\end{widetext}
$(\alpha, \beta)_{l} = \left[ e^{i\theta_{\beta,\alpha}} e^{i(\alpha-\beta)} - (-1)^{l} e^{-i\theta_{\beta,\alpha}} \right]/M \left[ e^{i(\beta -\alpha)}-1 \right]$, $l = 1,2$, 
where $\theta_{\beta,\alpha} = \theta(\beta) - \theta(\alpha)$,
$\theta(k)$ is defined by Eq.~(\ref{eq:theta}) and $\Delta_{1j}$ denotes removing the operators $G\left(\alpha_1\right)$ and $G\left(\alpha_j\right)$ from the matrix element that follows it. 

Eq.~(\ref{eq:strongcouplingWick}) can be solved analytically for $M \rightarrow \infty$ using the methods of Ref.~\citenum{Caianello2003} and numerically for finite systems as follows. Setting $n = 1$ in Eq.~(\ref{eq:strongcouplingWick}) and dividing both sides by $\bra{\Phi_{-}} \sigma_1^x \ket{ \Phi_{+}}$ results in a set of $M^2$ linear equations with complex coefficients
\begin{equation}\label{eq:MatrixEq}
\sum_{\alpha_1} (\alpha_1, \beta)_1 K \left( \alpha_1, \alpha_2 \right) = -(-\alpha_2, \beta)_2, 
\end{equation}
which can be grouped into $M$ sets (indexed by $\alpha_2$) with $M$ linear equations (indexed by $\beta$) in each set. Each set is equivalent to a matrix equation for a column of the matrix
\begin{equation}\label{eq:Ks}
K \left( \alpha_1, \alpha_2 \right) = \bra{\Phi_{-}} G\left(\alpha_2\right) G\left(\alpha_1\right) \sigma_1^x \ket{ \Phi_{+}} / \bra{\Phi_{-}} \sigma_1^x \ket{ \Phi_{+}}
\end{equation}
with fixed index $\alpha_2$, that can be solved by Gaussian elimination using an LU factorization~\cite{Press1992}. 

The denominator $\bra{\Phi_{-}} \sigma_1^x \ket{\Phi_{+}}$ in Eq.~(\ref{eq:Ks}) is determined by the completeness argument
\begin{equation}
\sum_{n = 0}^{N_{\textrm{max}}} \frac{1}{\left( 2n \right)!} \sum_{\left( \alpha \right)_{2n}} \left| \left\langle \Phi_- \left| G \left( \alpha_{2n} \right) ... G \left( \alpha_1 \right) \sigma_1^x \right| \Phi_+ \right\rangle \right|^2 = 1,
\label{eq:completeness_strong}
\end{equation}
where $N_{\textrm{max}} = M/2$ for even $M$ or $N_{\textrm{max}} = \left( M - 1 \right) / 2$ for odd $M$. The matrix elements in Eq.~(\ref{eq:completeness_strong}) can be recursively expanded using Eq.~(\ref{eq:strongcouplingWick}). We then substitute 
\begin{equation}
K \left( \alpha_1, \alpha_2 \right) = - i e^{i \left[ \theta \left( \alpha_1 \right) + \theta \left( \alpha_2 \right) \right] e^{-i \left( \alpha_1 + \alpha_2 \right)}} X \left( \alpha_1, \alpha_2\right),
\end{equation}
where $\theta \left( \alpha \right)$ is defined by Eq.~(\ref{eq:theta}) and $X \left( \alpha_1, \alpha_2\right)$ are real. The unimodular prefactors cancel out, and Eq.~(\ref{eq:completeness_strong}) reduces to 
\begin{equation}
\left| \left\langle \Phi_- \left| \sigma_1^x \right| \Phi_+ \right\rangle \right|^2 \sum_{n = 0}^{N_{\textrm{max}}} \frac{1}{\left( 2n \right)!} \sum_{\left( \alpha \right)_{2n}} \left[ \mathrm{Pf} \left( \left( \alpha \right)_{2n}  \right) \right]^2 = 1,
\end{equation}
where the Pfaffian satisfies the relation $\left[ \mathrm{Pf} \left( \left( \alpha \right)_{2n} \right) \right]^2 = \mathrm{det} \left[ X \left( \alpha_l , \alpha_m \right) \right],~1 \leq l,m \leq 2 n$~\cite{hochstadt1973}. Using a standard theorem (see Chapter 6 in Ref.~\citenum{hochstadt1973}), it now follows that 
\begin{equation}\label{eq:completeness}
1 = \left| \left\langle \Phi_{-} \left| \sigma_1^x \right| \Phi_{+} \right\rangle \right|^2 \det \left( I + X \right),
\end{equation}
where $I$ is the identity and $X$ is an $M \times M$ real matrix with elements $X \left( \alpha_l, \alpha_m\right)$. This expression can be evaluated by finding the eigenvalues of $X$. Since $X = -X^\dagger$, the spectrum of $X$ is purely imaginary. Let $Xu_j = i \lambda_j u_j$: then $Xu_j^* = -i \lambda_j u_j^*$, where $\lambda_j$ is real. When $M$ is odd, $\lambda_j =0$ is allowed.  Since the eigenvalues with $\lambda_j \neq 0$ come in complex conjugate pairs, for odd $M$ there will be an odd number of zero eigenvalues. Eq.~(\ref{eq:completeness}) reduces to:
\begin{equation}
1 = \left| \left\langle \Phi_- \left| \sigma_1^x \right| \Phi_+ \right\rangle \right|^2 \prod_{j=1}^{N_{\textrm{max}}} \left( 1 + \lambda_j^2 \right).
\end{equation}

Solution of Eqs.~(\ref{eq:MatrixEq}) and (\ref{eq:completeness}) completely defines the matrix element $\bra{\Phi_{-}} G\left(\alpha_2\right) G\left(\alpha_1\right) \sigma_1^x \ket{ \Phi_{+}}$. Higher-order matrix elements can then be calculated recursively, using the Pfaffian type solution~\cite{Caianello2003} generated by Eq.~(\ref{eq:strongcouplingWick}). The solution is thus an algorithm that relates $2n$-particle matrix elements to ones with $2n-2$ particles.  Iterating $n$ times yields a sum of products of $n$ contraction functions, resulting in a generalized form of Wick's theorem in which the in and out states are expressed naturally in terms of two different representations of the underlying Hilbert space.

The above procedure for evaluating dipole matrix elements between the eigenstates of Eq.~(\ref{eq:H_KMM}) is polynomial in the system size $M$, scaling as $M^3$.  This is the same as the complexity of diagonalization of the one-excitation subspace in the HL approximation, and is significantly better than the numerical effort for diagonalizing the complete KMM Hamiltonian, Eq.~(\ref{eq:H_KMM}) (which scales as $\propto 2^{3M}$)~\cite{Press1992}. To our knowledge this is the first use of such a technique for obtaining exact solutions of the generalized transition matrix elements $ \bra{\Phi_{-}} G\left(\alpha_{2n}\right) \dots G\left(\alpha_1\right) \sigma_1^x \ket{ \Phi_{+}}$ for finite system sizes. 

In the weak coupling regime, $B < 1$, it is necessary to calculate matrix elements between the ground state, $\left| \Phi_+ \right\rangle$, and states with an odd number of excitations. They can be expressed as
\begin{widetext}
\begin{equation}\label{eq:WeakMatrixEl}
\left\langle \Phi_{-} \left| G \left( \alpha_{2n+1} \right) \dots G \left( \alpha_1 \right) \sigma_1^x \right| \Phi_{+} \right\rangle =  
\sum_{j=1}^{2n+1} \left( -1 \right)^{(j-1)} K \left( \alpha_j \right) \Delta_j \left\langle \Phi_{-} \left| G \left( \alpha_{2n+1} \right) \dots G \left( \alpha_1 \right) \right| \Phi_{+} \right\rangle,
\end{equation}
\end{widetext}
where $\Delta_j$ denotes removing the operator $G \left( \alpha_j \right)$ from the matrix element that follows it. The one-particle function $K \left( \alpha_j \right) = \bra{\Phi_{-}} G(\alpha_j)\sigma_1^x \ket{ \Phi_{+}} / \langle \Phi_- | \Phi_+ \rangle$ in Eq.~(\ref{eq:WeakMatrixEl}) is constructed by setting $n = 0$ and is calculated from $K(\alpha_1,\alpha_2) = \langle \Phi_- | G(\alpha_2)G(\alpha_1) |\Phi_+\rangle/\langle \Phi_- | \Phi_+ \rangle$ as follows:

\begin{equation}
 K \left( \alpha_j \right) = M^{-1/2} \sum_{\alpha_0} e^{i \left[ \alpha_0 - \theta \left( \alpha_0 \right) \right]} \left[ K \left( \alpha_0 , \alpha_j \right) + \delta_{\alpha_0, -\alpha_j}\right].
\end{equation} 

Matrix elements of the type $\left\langle \Phi_{-} \left| G \left( \alpha_{2n} \right) \dots G \left( \alpha_1 \right) \right| \Phi_{+} \right\rangle$ can be expanded using Wick's theorem analogously to Eq.~(\ref{eq:strongcouplingWick}). However, they can also be obtained from matrix elements in the strong coupling regime by a duality argument. We introduce two unitary transformations, $D_{\pm} \Gamma_m D_{\pm}^\dagger = \Gamma_{m+1}$ for $1 \le m \le 2M-1$, $D_{\pm} \Gamma_{2M} D_{\pm}^\dagger = \mp \Gamma_1$: shifts with special boundary conditions. Here, spinors $\Gamma_{2m-1} = f_m^\dagger + f_m$ and $\Gamma_{2m} = -i \left( f_m^\dagger - f_m \right)$ satisfy the anti-commutation relations for a Clifford algebra: $\left[ \Gamma_n , \Gamma_m \right]_{+} = 2 \delta_{nm}$. Applying $D_{\pm}$ to $H_{\pm}$, we obtain $H'_{\pm} = D_{\pm} H_{\pm} D_{\pm}^\dagger$, where the transformed Hamiltonian is of KMM type, but with $b$ and $\varepsilon$ interchanged. The procedure for finding the ground state overlap $\left\langle \Phi_- \left| \right. \Phi_+ \right\rangle$ is analogous to that used for $\left\langle \Phi_{-} \left| \sigma_1^x \right| \Phi_{+} \right\rangle$ in the strong coupling regime.

\section{Linear Spectroscopy}
\label{sec:linearspectroscopy}

\begin{figure}[t]
\includegraphics[scale=1.0]{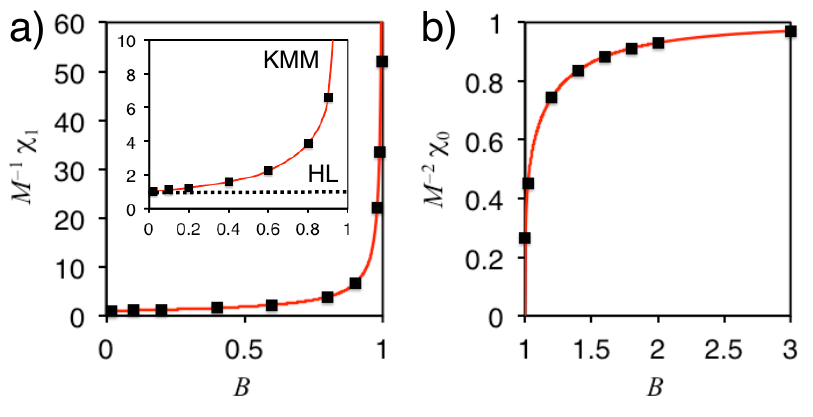}
\caption{(Color online) Total oscillator strength a) $\chi_1$ vs $B$ for excitation to the one-excitation manifold in weak coupling ($B<1$), and b) $\chi_0$ vs $B$ for excitation to the lowest excited state in strong coupling ($B>1$), for a system of size $M = 200$ (black squares) and $M \rightarrow \infty$ (red(gray) lines). The infinite size values are given by Eqs.~(\ref{eq:enhancementWeak}) and (\ref{eq:enhancementStrong}) for weak and strong coupling regimes, respectively. Inset: comparison of $\chi_1$ from KMM with values from HL.}
\label{fig:chi_values}
\end{figure}

This approach now allows a consistent calculation of the linear absorption spectra of dipole-coupled molecular arrays for any coupling strength $B$. We first consider the total oscillator strength $\chi_i$ for absorption from the ground state to the lowest excitation manifold, $i$. In the weak coupling regime, the lowest energy dipole-allowed transitions are from $\ket{\Phi_{+}}$ to one-excitation states and we have  \begin{equation}\label{eq:TotOscStrengthWeak}
\chi_1 = \sum_\alpha \left| \bra{\Phi_{-}} G \left( \alpha \right) \sum_{m=1}^M \sigma_m^x \ket{\Phi_{+}} \right|^2 .
\end{equation}
In the strong coupling regime, the lowest energy transition is from $\ket{\Phi_{+}}$ to $\ket{\Phi_{-}}$ and we have
\begin{equation}\label{eq:TotOscStrengthStrong}
\chi_0 =  \left| \bra{\Phi_{-}} \sum_{m=1}^M \sigma_m^x \ket{\Phi_{+}} \right|^2 .
\end{equation}
These expressions can be simplified using the translational symmetry of Eq.~(\ref{eq:H_KMM}). By translational invariance, Eq.~(\ref{eq:TotOscStrengthWeak}) reduces to 
\begin{equation}\label{eq:TotOscStrengthStrongSimplified}
\chi_1 = M^2 \left| \left\langle \Phi_- \left| G \left( 0 \right) \sigma_1^x \right| \Phi_+ \right\rangle \right|^2.
\end{equation}
From Eqs.~(\ref{eq:BogoliubovValatin}), (\ref{eq:theta}), and (\ref{eq:excenergy}) it follows that the energy in the $\alpha = 0$ mode is $E_0 \left( 0 \right) F^\dagger \left( 0 \right) F \left( 0 \right)$. To satisfy Eq.~(\ref{eq:TotOscStrengthStrong}), as $B$ increases through unity, the new ground state of $H_{-}$, i.e., the lowest excited state for $B > 1$, must incorporate an $\alpha = 0$ excitation, $G^\dagger \left( 0 \right) = F^\dagger \left( 0 \right)$.
The right hand side of Eq.~(\ref{eq:TotOscStrengthStrongSimplified}) then goes smoothly to 
\begin{equation}
\chi_0 = M^2 \left| \left\langle \Phi_- \left| \sigma_1^x \right| \Phi_+ \right\rangle \right|^2.
\end{equation}
Thus, the oscillator strength $\chi_0$ for $B > 1$ scales quadratically with $M$. The linear $M$ scaling for $B < 1$ is a consequence of the correct scaling of the matrix element which absorbs a single power of $M$. 

Eqs.~(\ref{eq:TotOscStrengthWeak}) and (\ref{eq:TotOscStrengthStrong}) can be explicitly evaluated using the matrix elements for finite $M$ derived above. The results are shown by black squares in Figure~\ref{fig:chi_values}. Solutions for $M \longrightarrow \infty$ may be obtained using the analytic methods of Ref.~\citenum{Abraham2012}, which yields 
\begin{equation}\label{eq:enhancementWeak}
A \left[ B \right] =\lim_{M \rightarrow \infty}  M^{-1}\chi_1 = \left( 1 -  B \right)^{-\frac{3}{4}} \left( 1 + B \right)^{\frac{1}{4}}
\end{equation}
and
\begin{equation}\label{eq:enhancementStrong}
\widetilde{A} \left[ B \right] = \lim_{M \rightarrow \infty}  M^{-2} \chi_0 = \left(1 - 1/B^{2} \right)^{\frac{1}{4}}.
\end{equation}
These solutions reveal the size scalings $\chi_1 \propto M$, $\chi_0 \propto M^2$ and are plotted as red(gray) lines in Figure~\ref{fig:chi_values}. We find excellent agreement with the finite-size values for $M = 200$ everywhere except very close to $B=1$, where in the infinite size limit $M^{-1} \chi_1$ diverges and $M^{-2} \chi_0$ goes to 0 (see Appendix~\ref{sec:totaloscstrengthandabsdens}).

Our analysis shows that only the lowest excited state, $\alpha=0$, contributes to the oscillator strength $\chi_1$, while $\chi_0$ is determined by the single transition from $\ket{\Phi_+}$ to $\ket{\Phi_-}$. The inset in Figure~\ref{fig:chi_values}a shows that in the weak coupling regime the absorption to and hence the emission from the lowest excited state scales linearly with $M$, for both finite $M$ and  $M\rightarrow \infty$. This is consistent with a one-photon superradiance~\cite{dicke1954coherence}.  Superradiance with linear scaling is also seen in the HL limit, with a pre-factor $A_{\mathrm{HL}} \equiv 1$ that is independent of coupling strength $B$~\cite{note_HLsuperradiance}. However $\chi_1$ has a pre-factor that is equal to the HL value only when $B$ is extremely small and that increases with $B$, indicating an excess superradiance. This excess superradiance diverges for $M \rightarrow \infty$ as the critical point at $B=1$ is approached from below.

The strong coupling regime shows an even more interesting superradiant behavior, since here the excited state $\ket{\Phi_{-}}$ is superradiant with a rate $\propto M^2$, and $M^{-2}\chi_0$ is asymptotic to $1$ for $B\rightarrow\infty$. This anomalous scaling does not correspond to that of a one-photon superradiance, but rather to the scaling normally associated with the maximum superradiance possible in an ensemble of two-level systems where all two-level systems are excited~\cite{dicke1954coherence}. Thus, it constitutes a more radical enhancement of one-photon superradiance than the pre-factor enhancement seen in the weak coupling regime. The latter can also be engineered for non-interacting two-level systems by making use of conditioned state preparation~\cite{scully2009super} and has been termed ``super superradiance''. Given the dipolar interactions in the array, it is perhaps not surprising to find enhanced pre factors for one-photon superradiance, or ``super superradiance".  The increasing enhancement with $B$ and its divergence as $B$ approaches unity reflects the change in nature of the eigenstates as the transition to the (anti)ferroelectric ordered phase is approached. However, beyond this transition, in the ordered phase, we find the remarkable result that a single excitation can give rise to a superradiance that shows the scaling characteristic of a non-interacting system with $M$ excitations. This constitutes a one-photon analog of hyperradiance~\cite{gronbech1993hyperradiance} and reflects the radical change in nature of the eigenstates on going from the disordered paraelectric phase, $B<1$, to the (anti)ferroelectric states for $B > 1$.

We already specified that in the limit $M \rightarrow \infty$, $M^{-1} \chi_1$ diverges as $B \rightarrow 1-$. It is informative to analyze this quantity as a fluctuation sum of pair correlations of transition dipole moments, $C(m-n) = \sum_\alpha \bra{\Phi_+} \sigma_m^x G^\dagger(\alpha) \ket{\Phi_-} \bra{\Phi_-} G(\alpha) \sigma_n^x \ket{\Phi_+}$ ($B <1$), for which the correlation propagates solely through single excitations. Carrying out the sum over $\alpha$ and then using translational symmetry, we find 
\begin{equation}
\lim_{M \rightarrow \infty} M^{-1}\chi_1 = C(0)+2 \sum_{m=1}^{\infty} C(m),
\end{equation}
with
\begin{equation}
C(m) = \frac{(1-B^2)^{1/4}}{2\pi} \int_0^{2\pi} dk \frac{e^{imk}}{(1+B^2-2B\cos k)^{1/2}}.
\end{equation}
Evaluation of the integral for large $m$ reveals that the correlations decay on a length scale $\left( 1 - B \right)^{-1}$, which diverges at the quantum critical point $B = 1$. In contrast, for the HL approximation, $C(m)=\delta_{m0}$. Thus not only does HL underestimate the oscillator strength and hence the extent of superradiance for $B < 1$, it also shows no divergence at the critical point (see inset in Figure~\ref{fig:chi_values}a). HL is furthermore inapplicable in the strong coupling regime, where it gives an incorrect energy spectrum.

\begin{figure}[t]
\includegraphics[scale=1.0]{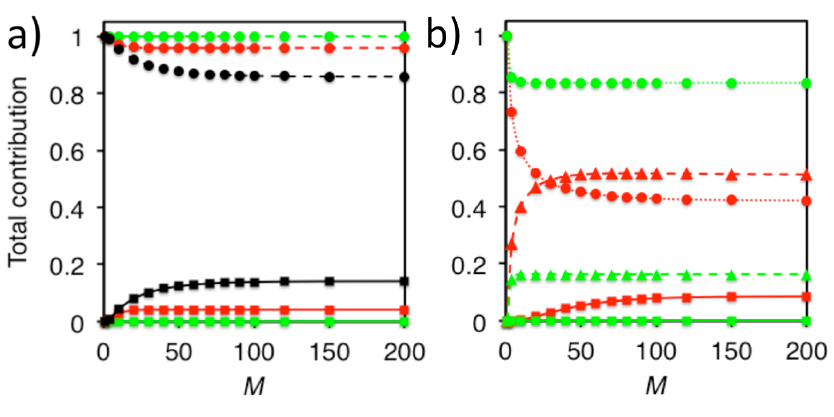}
\caption{(Color online) Total contribution (per molecule) of manifolds with different numbers of excitations to the total oscillator strength from the ground state $\ket{\Phi_+}$. (a) Weak coupling regime: $B = 0.4$ (green(light gray)), $0.9$ (red(dark gray)), $0.98$ (black). Dashed(solid) lines represent the contribution of the one(three)-excitation manifolds. (b) Strong coupling regime: $B = 1.02$ (red(dark gray)) and $1.4$ (green(light gray)). Dotted lines represent the contribution of the $\left| \Phi_- \right\rangle \longrightarrow \left| \Phi_+ \right\rangle$ transition, dashed(solid) lines -- the contribution of excitations from the ground state $\ket{\Phi_+}$ to the two(four)-excitation manifold. Points represent the calculated values for specific system size $M$.}
\label{fig:contributions}
\end{figure}

\begin{figure*}[t]
\includegraphics[scale=1.0]{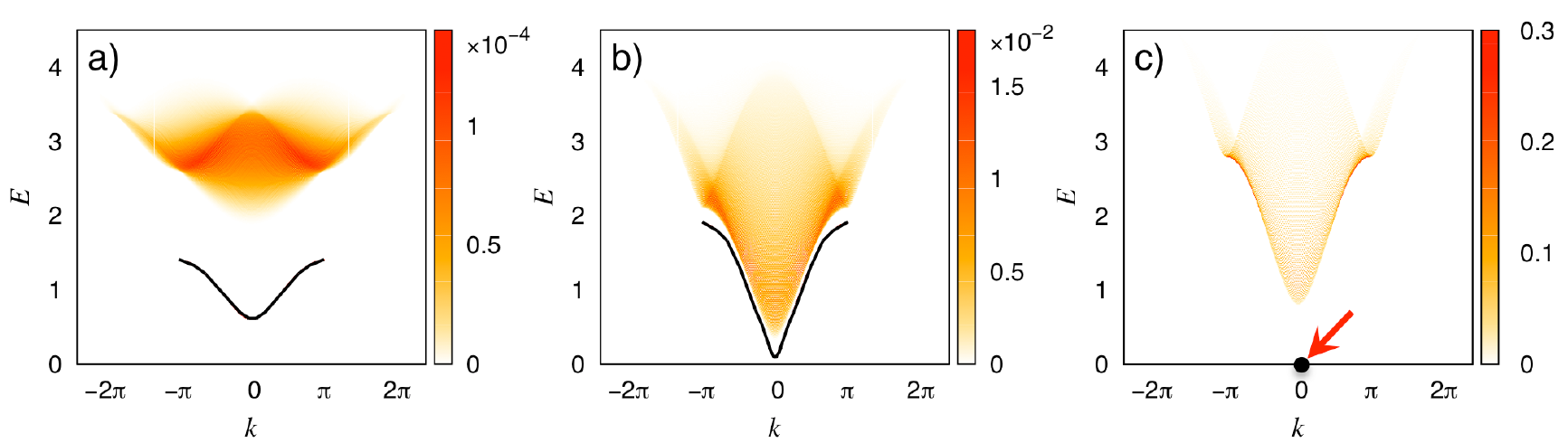}
\caption{(Color online) Linear absorption density $\rho_A \left( k, E \right)$ from the ground state $\ket{\Phi_+}$ of an array with $M=200$: to the three-excitation manifold, $G^\dagger \left( \alpha_1 \right) G^\dagger \left( \alpha_2 \right) G^\dagger \left( \alpha_3 \right) \ket{\Phi_{-}}$, for $B=$ 0.40 (a) and 0.90 (b); to the two-excitation manifold, $G^\dagger \left( \alpha_1 \right) G^\dagger \left( \alpha_2 \right) \ket{\Phi_{-}}$, for $B = 1.40$ (c). Black lines in panels (a) and (b) indicate the dispersion curve of the one-excitation manifold, $G^\dagger \left( \alpha_1 \right) \ket{\Phi_{-}}$. In the strong coupling regime the lowest excitation manifold collapses to a single transition 
$\ket{\Phi_{+}} \longrightarrow \ket{\Phi_{-}}$, indicated in panel (c) by the black dot and red(gray) arrow. Energy is in units of $\varepsilon$; wavenumber is in radians per lattice constant of the chromophore array. Calculations were performed on a $639 \times 480$ grid over $k \in \left[-10,10\right)$ $\times E \in \left[ 0,6 \right)$.}
\label{fig:multiexc}
\end{figure*}

Another unusual aspect of linear spectroscopy with the KMM eigenstates is the presence of finite oscillator strength from the ground state to manifolds of states with multiple excitations. As explained above and also noted in earlier work focused on the extremely weak coupling limit, $B \ll 1$~\cite{bakalis1997optical,agranovich2000frenkel}, such excitations are not allowed in the HL description and are a signature of the double excitation and de-excitation terms, $P^\dagger_m P^\dagger_{m+1}$ and $P_m P_{m+1}$, in Eq.~(\ref{eq:H_KMM}). Our spectroscopic analysis allows one to extract the contribution of a manifold with any given number of excitations to the total oscillator strength for arbitrary $B$. These contributions (per molecule) are shown in Figure~\ref{fig:contributions} as a function of the system size $M$.  In both weak and strong coupling regimes, the contributions of the higher excitation manifolds are most significant close to the critical point $B = 1$. We note however, that even at $B = 0.98$ and $B = 1.02$, the contributions of all excitation manifolds beyond the third or fourth, respectively, are negligible. The $M\rightarrow \infty$ limit analysis~\cite{Abraham2012} of oscillator strengths from the ground state $\ket{\Phi_+}$ to higher excitation number manifolds, i.e., $\chi_{2 n + 1}$ (weak coupling) and $\chi_{2 n}$ (strong coupling) can be shown to possess the same linear $M$ scaling and critical exponent $-3/4$ as $\chi_1$. This agrees well with the finite size calculation results for larger $M$ values shown in Figure~\ref{fig:contributions}.

\begin{figure}[b]
\includegraphics[scale=1.0]{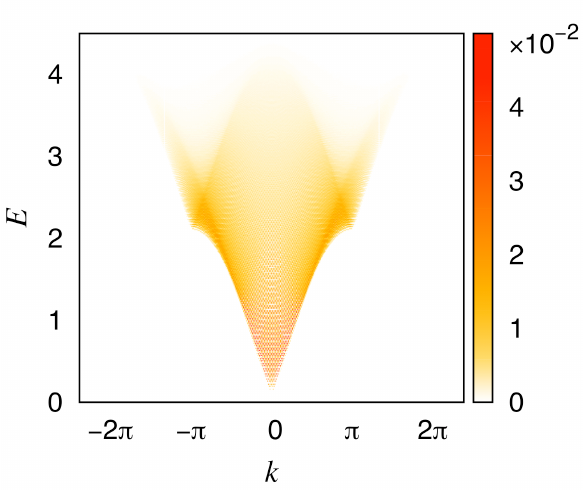}
\caption{(Color online) Linear absorption density $\rho_A \left( k, E \right)$ from the ground state $\ket{\Phi_+}$ of an array with $M=200$ to the four-excitation manifold, $G^\dagger \left( \alpha_1 \right) G^\dagger \left( \alpha_2 \right) G^\dagger \left( \alpha_3 \right) G^\dagger \left( \alpha_4 \right) \ket{\Phi_{-}}$, for $B=$ 1.02. Energy is in units of $\varepsilon$; wavenumber is in radians per lattice constant of the chromophore array. Calculations were performed on a $639 \times 480$ grid over $k \in \left[-10,10\right)$ $\times E \in \left[ 0,6 \right)$.}
\label{fig:fourexc}
\end{figure}

Since the number of states in higher excitation manifolds is large (e.g., 19900(1313400) two(three)-excitation states for $M=200$), we sum over individual transitions in a given $k$ and $E$ interval to obtain a linear absorption density per unit momentum transfer and energy, $\rho_A \left( k, E \right)$, which displays the key features of the multi-excitation transitions. The absorption density is defined as $\rho_A \left( k, E \right) = \sum \left| \mu_n(k',E') \right|^2 \delta k^{-1} \delta E^{-1}$, where $\mu_n(k',E')$ is a transition matrix element to a state in the $n$-excitation manifold with total momentum $k'$ and total energy $E'$, and the summation is over all states in that manifold with $k' \in \left[ k, k+\delta k\right)$, $E' \in \left[ E, E+\delta E \right)$. The intervals $\delta k$ and $\delta E$ define the momentum and energy resolution, respectively, of Figures~\ref{fig:multiexc} and \ref{fig:fourexc}. Figure \ref{fig:multiexc} shows $\rho_A \left( k, E \right)$ from the ground state to the three-excitation manifold ($B< 1$) and two-excitation manifold ($B> 1$) for an array of $M=200$ chromophores. It is evident from Figure~\ref{fig:contributions} that the absorption density for the four-excitation manifold is only appreciable for $B \longrightarrow 1+$. Figure~\ref{fig:fourexc} shows the absorption density to this manifold for $B = 1.02$. Note, however, that even at this $B$ value the maximum absorption density for the two-excitation manifold is three orders of magnitude higher than for the four-excitation manifold (see Appendix~\ref{sec:totaloscstrengthandabsdens}). Absorption densities for all excitation manifolds beyond the fourth are very small.

Just as for absorption to the single-excitation manifold, the single-photon absorption to multi-excitation manifolds is very different in the strong and weak coupling regimes. In the weak coupling regime (Figure \ref{fig:multiexc}a,b) the absorption density from the ground state, $\ket{\Phi_+}$, to three-excitation states, $G^\dagger \left( \alpha_3 \right) G^\dagger \left( \alpha_2 \right) G^\dagger \left( \alpha_1 \right) \ket{\Phi_{-}}$, increases with $B$ (note the different range of the color bar scale for panels (a) and (b)). While the transitions with maximum oscillator strengths are always located at $k = 0$ for $b < 0$, the maximum value of $\rho_A \left( k, E \right)$ is nevertheless located close to $k = \pm \pi$ as a result of the higher density of states  there. At $B =1$, the parity of the eigenstates of $H_{-}$ changes (see Section~\ref{sec:reviewofeigenstatecalculation}), so that in the strong coupling regime transitions from $\ket{\Phi_{+}}$ to the two-excitation manifold, $G^\dagger \left( \alpha_1 \right) G^\dagger \left( \alpha_2 \right) \ket{\Phi_{-}}$, are now allowed (Figure~\ref{fig:multiexc}c). In contrast to the weak coupling regime, as $B$ increases beyond unity, transitions to the higher-excitation manifolds are increasingly suppressed until $\ket{\Phi_{+}} \rightarrow \ket{\Phi_{-}}$ becomes the only allowed transition and saturates the oscillator strength. Because of the asymptotic degeneracy of $\ket{\Phi_{\pm}}$ for $M\rightarrow \infty$, the energy of this transition decreases to zero as $M$ increases, implying strong absorption for arbitrarily small $E_- - E_+$.

\section{Realization with trapped dipolar molecules}
\label{sec:realization}

The spectral features predicted in Section~\ref{sec:linearspectroscopy} can be observed by emulation with currently available quantum technology. The transverse Ising Hamiltonian of Eq. (\ref{eq:ising}) has already been implemented for a broad range of $B$ values spanning both weak and strong coupling regimes in experiments using finite-size chains of trapped ions~\cite{Kim2011,Islam03052013,Senko25072014, islam2011onset} and simulations in the strongly-correlated phase $B>1$ have also been made with neutral atoms in optical lattices~\cite{Simon2011}. Theoretical studies have shown that a variety of related spin-lattice Hamiltonians can also be simulated using the ground and first excited rovibrational states of the ground electronic potential of polar diatomic molecules trapped in an optical lattice~\cite{Micheli2006,gorshkov2011quantum,Gorshkov2011} and recently the $XY$ lattice spin model in a three-dimensional lattice was implemented using cold KRb molecules~\cite{yan2013observation}. 

Here we describe a possible implementation of Eq.~(\ref{eq:H_KMM}) that utilizes $^2\Sigma$ ground state molecules in an optical lattice, together with a static magnetic field, an off-resonant near IR continuous wave (cw) laser field, and two near-resonant cw fields~\cite{herrera2014entanglement}. In such a system the Hamiltonian is of the form
\begin{equation}
H = \sum_i H_i + \sum_{j < i} V_\mathrm{dd} \left( \mathbf{R}_{ij}, \theta_i, \phi_i, \theta_j, \phi_j \right) ,
\end{equation}
where $\mathbf{R}_{ij}$ is the intermolecular separation vector,  $\theta_i$ and $\phi_i$ are the azimuthal and polar angles of molecule $i$ with respect to the spin quantization axis, and $H_i$ is
the single molecule Hamiltonian. The latter is of the form
\begin{multline}
H_i = B_e \mathbf{N}_i^2 + \gamma_\mathrm{sr} \mathbf{N}_i \cdot \mathbf{S}_i + g_S \mu_B B_0 S_{z,i} \\ -  \frac{\Delta \alpha |E_0|^2}{4} C_2^0 \left( \theta_i \right) \otimes I_s ,
\end{multline}
where $B_e$ is the rotational constant, $\mathbf{N}$ is the rotational angular momentum, $\gamma_\mathrm{sr}$ is the spin-rotation constant, $\mathbf{S}$ and $S_z$ are the electron spin angular momentum and its projection along the $z$-axis, $g_S \approx 2$ is the electron spin g-factor, $\mu_B$ is the Bohr magneton, $B_0 \mathbf{z}$ is the applied magnetic field, $\Delta \alpha$ is the polarizability anisotropy, $E_0$ the applied cw field amplitude, $C_2^0 \left( \theta_i \right) = \left( 3\cos^2 \theta_i - 1 \right)/2$ is a Racah normalized spherical harmonic, and $I_s$ is the identity in the electron spin subspace. Pairs of molecules will interact via the electric dipole-dipole interaction
\begin{equation}
V_\mathrm{dd} = \frac{d^2}{r_{ij}^3} \left( 1- 3 \cos^2 \Theta \right) d_0^i d_0^j ,
\end{equation}
where $d$ is the electric dipole moment in the molecule-fixed frame, $r_{ij} = |\mathbf{R}_{ij}|$ is the intermolecular distance, $\Theta$ is the (fixed) angle between the spin quantization axis and the intermolecular separation vector, and $d_q^i$ ($q = -1, 0, 1$) is the dimensionless dipole moment operator.

Defining the states $\ket{g} = \ket{N = 0, M_N = 0}\ket{\uparrow}$ and $\ket{g'} = \ket{N = 0, M_N = 0}\ket{\downarrow}$, where $\ket{N, M_N}$ is an eigenstate of the operators $\mathbf{N}$ and $N_z$, we introduce the following states~\cite{herrera2014entanglement}:
\begin{equation}\label{eq:ground}
\ket{D} = \cos \phi \ket{g} - \sin \phi \ket{g'}
\end{equation}
and
\begin{multline}\label{eq:excited}
\ket{e} = \sqrt{1-a} \ket{N = 1, M_N = 0}\ket{\downarrow} \\
- \sqrt{a} \ket{N = 1, M_N = -1}\ket{\uparrow},
\end{multline}
where $a \approx \eta^2/2$, $\eta = \gamma_\mathrm{sr}/g_S \mu_B B_0$, and $\phi$ is the mixing angle between the high- and low-field seeking states $\ket{g'}$ and $\ket{g}$, respectively, in the absence of the cw laser fields. 

We utilize the states given by Eqs.~(\ref{eq:ground}) and (\ref{eq:excited}), prepared according to the procedure described in Ref.~\citenum{herrera2014entanglement}, as our two-level system and identify the excitation energy $\varepsilon$ of Eq.~(\ref{eq:H_KMM})
with the single molecule energy difference $\epsilon_e$ between the states $\ket{D}$ and $\ket{e}$.  In this basis, the dipole-dipole interaction, $V_\mathrm{dd}$, is given by
\begin{equation}
V_{ij} = b_{ij} \left[\ketbra{e_i e_j}{D_i D_j} + \ketbra{e_i D_j}{D_i e_j} + \mathrm{h.c.}\right]
\end{equation}
with
\begin{equation}\label{eq:coupling}
b_{ij} = \frac{1}{3} \frac{d^2}{r_{ij}^3}(1 - 3 \cos^2 \Theta)(1 - \eta^2)(1 - \delta^2),
\end{equation}
where $\delta = |\pi/2 - \phi|$, subject to $\eta \ll 1$ and $\delta \ll 1$.

Implementing the Hamiltonian of Eq.~(\ref{eq:H_KMM}) using open-shell diatomic polar molecules also allows tuning the site energy $\varepsilon$ to values well below the strength of the nearest-neighbour dipole-dipole interaction $b_{ij}\equiv b$~\cite{herrera2014entanglement}. This makes it possible to prepare a one-dimensional molecular array with $B$ spanning the entire range of values from $B \ll 1$ in the extreme weak coupling regime, through the critical point at $B=1$, to deep in the strongly-correlated phase $B \gg 1$. 
We note that in a realistic optical lattice, the array size $M$ is always small enough to keep the energy gap $E_{-} - E_{+}$ (see Eq.~(\ref{eq:vactovacexcen})) finite for all achievable values of $B\gg 1$. Energy scales on the order of hertz can be resolved spectroscopically~\cite{ospelkaus2010controlling} and typical values of $b_{ij}$ are tens of kHz, so $E_{-} - E_{+}$ needs only to be reduced to within this order of magnitude in order to see the effects predicted in Section~\ref{sec:linearspectroscopy}. 
Moreover, currently available nanoplasmonic lattices~\cite{gullans2012nanoplasmonic} on a chip have opened the possibility to reduce the intermolecular separation distance by an order of magnitude compared with optical lattices, which would enhance the interaction energy $b_{ij}$ by a factor of $10^3$. Under these conditions, the spectral gap $E_{-} - E_{+}$ in the regime $B\gg 1$ is on the order of MHz for finite arrays and thus still readily accessible with radio frequencies.

The intrinsically slow radiative decay of the rotational state $\ket{e}$ may be overcome and the hyperradiant emission from the array detected by mixing in a small amount of a short-lived electronically excited rotational state $\ket{f}$ into the single molecule states $\ket{e}$. This can be achieved by irradiation with a weak cw laser to adiabatically form the single-molecule superposition state $\ket{\psi} = \sqrt{1-x}\ket{e}-\sqrt{x}\ket{f}$ with mixing coefficient $0 < x\ll 1$. When a weakly allowed transition $\ket{e} \leftrightarrow \ket{f}$ is then combined with a fast decay $\ket{f} \rightarrow \ket{D}$, the one-excitation decay rate of the superposition state $\ket{\psi}$ for large $M$ can be calculated as follows. In the weak coupling regime, $\gamma_M = A \left[ B \right] M x \gamma_f$, where $A \left[ B \right] \geq 1$ is given by Eq.~(\ref{eq:enhancementWeak}). We find that $A \left[ B \right]$ increases with $B$, significantly exceeding the constant value $A_{\mathrm{HL}}[B] \equiv 1$ that is predicted by the HL approximation as $B = 1$ is approached: e.g., $A[0.98] \approx 22$. In the strong coupling regime, the one-excitation decay rate scales quadratically with the size of the system: $\gamma_M = \widetilde{A} \left[ B \right] M^2 x \gamma_f$, where $\widetilde{A} \left[ B \right] \leq 1$ is given by Eq.~(\ref{eq:enhancementStrong}). We find $\widetilde{A} \left[ 1.02 \right] \approx 0.44$, and $\widetilde{A} \left[ 1.4 \right] \approx 0.84$. If $M = 200$, this corresponds to enhancements by a factor of 88 for $B = 1.02$ and a factor of 168 for $B = 1.4$ relative to the $B$-independent HL emission rate in the weak coupling regime. For alkaline earth monohalides, typical electronic excited state decay rates are $\gamma_f\sim 10 -50$ MHz~\cite{Dagdigian1974}. This yields radiative lifetimes $1/\gamma_M\sim 0.05-0.25$ ns for $x=0.1$, $M = 200$, and $B = 0.98$ and $\sim 0.005-0.03$ ns for $B = 1.4$ (with the same $x$ and $M$), implying that such an emulator can emit with a size-enhanced rate at optical frequencies.

Such experiments will be interesting for investigation of the effects of molecular dipole interactions coupling sites beyond nearest neighbors, which are not included in Eq.~(\ref{eq:H_KMM}). The fact that the dipolar interaction is formally short-ranged in one dimensional systems~\cite{dutta2001phase} means the critical properties are very similar to those of a nearest neighbor dipolar-coupled model~\cite{deng2005}. We may therefore expect that the main $B$-dependent features of the spectroscopy predicted from Eq.~(\ref{eq:H_KMM}) will be maintained in the presence of beyond nearest neighbor interactions. Future work will address the detailed effects of such longer range interactions.

\section{Summary and conclusions}
\label{sec:summary}

The current paper presents a consistent study of linear spectroscopy for both infinite and finite arrays of dipole-coupled two-level molecules, described by the Hamiltonian given by Eq.~(\ref{eq:H_KMM}). We propose a new exact method of calculating transition matrix elements for finite numbers of molecules $M$ and arbitrary coupling strength $B$ between molecules in an array. We also perform analytical calculations of transition matrix elements for $M \longrightarrow \infty$.

Our analysis reveals distinct spectroscopic signatures of weak ($B < 1$) and strong ($B > 1$) coupling regimes, separated for infinite size arrays by a quantum critical point. This is a consequence of the different many-body nature of energy eigenstates for the two regimes: for $B < 1$, the ground state of the molecular array is disordered, but for $B > 1$ it has (anti)ferroelectric ordering. Direct optical transitions from the ground state to states with multiple molecular excitations (odd excitation numbers for $0 < B < 1$ and even excitation numbers for $B > 1$) are permitted. We analyze the scaling of absorption and emission with system size and find that the oscillator strengths show enhanced superradiant behavior in both ordered and disordered phases. As the coupling increases, the single excitation oscillator strength rapidly exceeds the well known Heitler-London value. For $M \longrightarrow \infty$, we find a novel, singular spectroscopic signature of the quantum phase transition between disordered and ordered ground states. In the strong coupling regime we show the existence of a unique spectral transition with excitation energy that can be tuned by varying the system size and that asymptotically approaches zero for $M\rightarrow \infty$. The oscillator strength for this transition scales quadratically with system size. The change from linear to quadratic scaling of one-photon absorption and emission as the coupling strength is increased beyond the quantum critical point at $B = 1$ constitutes a one-photon analog of the anomalous size scaling of superradiance, termed hyperradiance, that is seen in phase-locked soliton oscillators~\cite{gronbech1993hyperradiance}. Finally, we show how arrays of ultra cold dipolar molecules trapped in an optical lattice can be used in a quantum emulation of Eq.~(\ref{eq:H_KMM}) to access the strong coupling regime and observe the anomalous superradiant effects associated with this regime. 

The theoretical approach presented here may be readily extended and applied to the analysis of non-linear spectroscopy for dipole-coupled non-polar chromophore arrays with arbitrary coupling strength~\cite{tobepublished}.

\appendix
\section{Translational Symmetry}
\label{sec:translationalsymmetry}

Since all molecules in the arrays that we study are identical and are coupled to their neighbors in the same way, the Hamiltonian given by Eq.~(\ref{eq:H_KMM}) has translational symmetry when periodic boundary conditions are assumed. The translation operator $T$ is such that $TP_mT^\dagger = P_{m-1}$, $2 \le m \le M$ with $TP_1T^\dagger = P_M$. Using this translational symmetry and the definition of $F^\dagger \left( k \right)$, Eq.~(\ref{eq:Fourier}), the following result may be derived \cite{Abraham1967}: if $\exp \left( iMk_j \right) = \left( -1 \right)^{n-1}$ for 
$1 \leq j \leq M$, then
\begin{equation}
\label{eq:trans_symm}
\begin{split}
& T F^\dagger \left( k_1 \right) ... F^\dagger \left( k_n \right) \left| 0 \right\rangle = \\ & = \exp\left( i \sum_{j=1}^n k_j \right) F^\dagger \left( k_1 \right) ... F^\dagger \left( k_n \right) \left| 0 \right\rangle.
\end{split}
\end{equation}
This is the origin of the rather curious periodic and anti-periodic wavenumbers that arise in the analysis of the KMM Hamiltonian. Applying Eq.~(\ref{eq:trans_symm}), we see that $T \left|\Phi_+ \right\rangle = \left| \Phi_+ \right\rangle$ and that for the strong coupling regime, $B=2 \left| b \right|/\varepsilon > 1$, $T \left|\Phi_- \right\rangle = \left| \Phi_- \right\rangle$, where $\left| \Phi_- \right\rangle$ is an eigenvector of both the Hamiltonian $H$ and of $H_-$. However, when $B < 1$, $T \left|\Phi_- \right\rangle \neq \left| \Phi_- \right\rangle$, and $\left| \Phi_- \right\rangle$ is an eigenvector of $H_-$, but \textit{not} of $H$. In this case, $G^\dagger \left( \alpha \right) \left| \Phi_- \right\rangle$ is an eigenvector of $H$ and of $T$ with eigenvalue $\exp \left( i \alpha \right)$. For the 3-particle states, we have
\begin{equation}
\begin{split}
& T G^\dagger \left( \alpha_1 \right) G^\dagger \left( \alpha_2 \right) G^\dagger \left( \alpha_3 \right)  \left| \Phi_- \right\rangle = \\ & = \exp \left[ i \left( \alpha_1 + \alpha_2 + \alpha_3 \right) \right] G^\dagger \left( \alpha_1 \right) G^\dagger \left( \alpha_2 \right) G^\dagger \left( \alpha_3 \right) \left| \Phi_- \right\rangle.
\end{split}
\end{equation}

\section{Lowest Excitation Energy in the Strong Coupling Regime}
\label{sec:lowestexceninstrongcoup}

The energy of the lowest excitation, $E_{-} - E_{+}$, can be calculated using the Cauchy integral formula with appropriate kernels to implement the summations in Eq.~(\ref{eq:lowestexcen}). This procedure uses the properties that $E \left( k \right) = E \left( k + 2 \pi \right)$ and that $E \left( k \right)$ is analytic for $\left| \mathrm{Im}~k \right| < v_0$, with $\cosh v_0 = \left( B + B^{-1} \right) / 2$. After some transformations, we arrive at:
\begin{equation}\label{eq:intvactovacexcen}
E_{-}-E{+} = \frac{M}{\pi} \int_{-\pi + i \varepsilon}^{\pi+i \varepsilon} \mathrm{d}k \frac{E \left( k \right) \exp \left( i M k \right)}{1 - \exp \left( 2 i M k \right)}.
\end{equation}
Here $E \left( k \right)$ has a branch point at $k = \pm i v_0$ and the plane may cut along $\left( i v_0 , \infty \right)$ and its mirror image in the real axis ($ \mathrm{mod}~2 \pi $). Deforming the contour for the line integral in Eq.~(\ref{eq:intvactovacexcen}) into a ``hair pin'' on the upper half plane cut, where $E \left( k \right)$ is purely imaginary and reverses sign on crossing the cut, leads to Eq.~(\ref{eq:vactovacexcen}), a useful integral representation.

\section{Total Oscillator Strength to the Lowest Excitation Manifold and Absorption Densities for Transitions to Higher Excitation Manifolds}
\label{sec:totaloscstrengthandabsdens}

\begin{figure}[t]
\begin{center}
\includegraphics[scale=1.0]{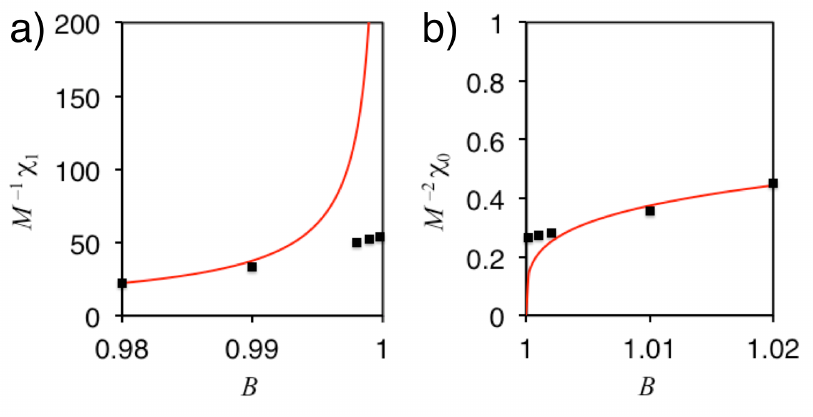}
\end{center}
\caption{Total oscillator strength close to the critical point $B = 1$: a) $\chi_1$ vs $B$ for excitation to the one-excitation manifold in weak coupling ($B < 1$), and b) $\chi_0$ vs $B$ for excitation  to the lowest excited state in strong coupling ($B > 1$), for a system of size $M = 200$ (black squares) and $M \longrightarrow \infty$ (red(gray) lines).}
\label{fig:partsuscept01}
\end{figure}

The total oscillator strength close to the critical point $B = 1$ is shown in Figure~\ref{fig:partsuscept01}. It shows a divergence for $M \longrightarrow \infty$, but not for finite $M$ values.

\begin{figure*}[b]
\begin{center}
\includegraphics[scale=0.9]{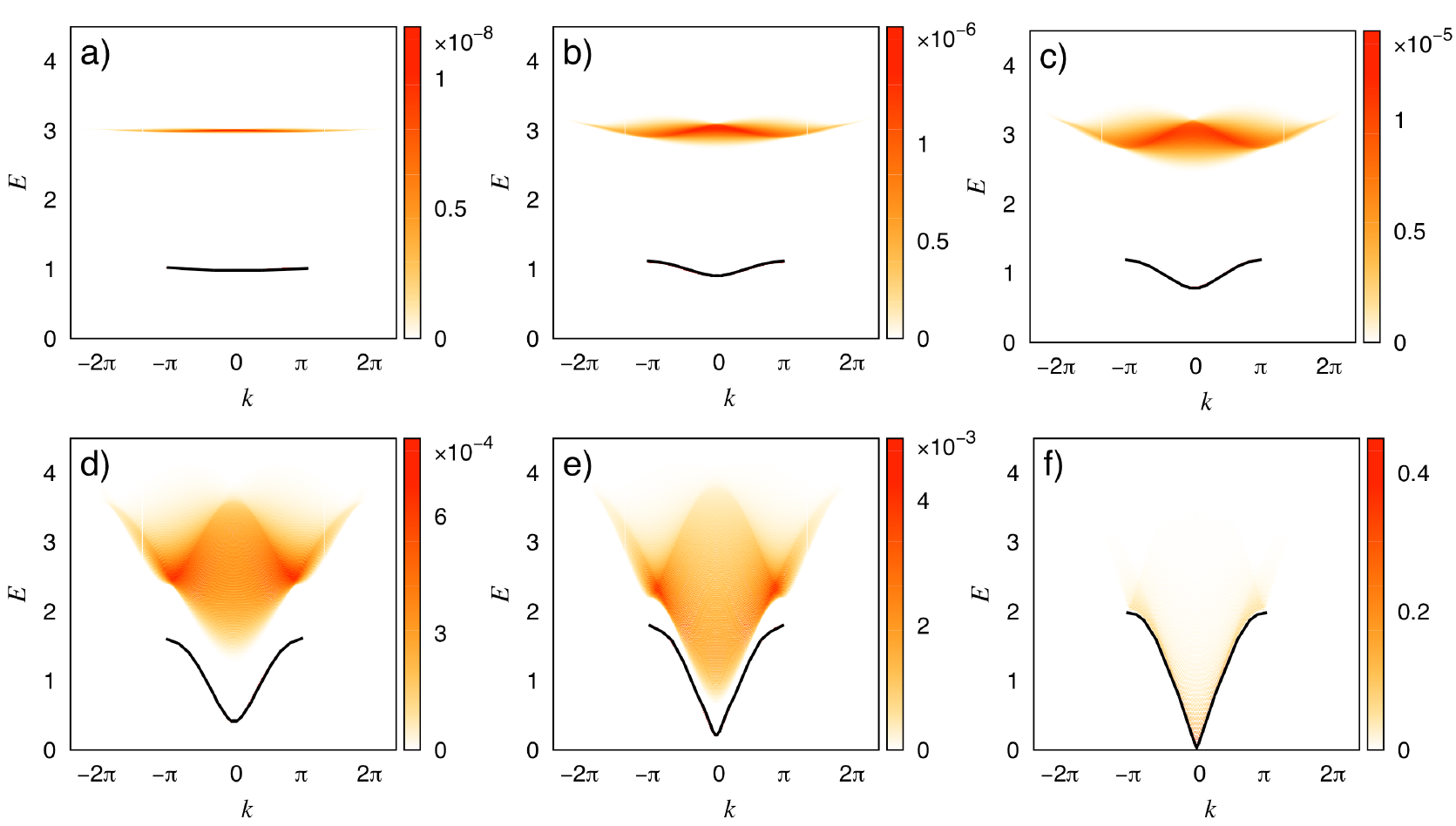}
\end{center}
\caption{Linear absorption density $\rho_A \left( k, E \right)$ from the ground state $\left| \Phi_+ \right\rangle$ of an array with $M=200$ to the three-excitation manifold, $G^\dagger \left( \alpha_1 \right) G^\dagger \left( \alpha_2 \right) G^\dagger \left( \alpha_3 \right) \left| \Phi_{-} \right\rangle$, for $B=$ 0.02~(a), 0.1~(b), 0.2~(c), 0.6~(d), 0.8~(e), and 0.98~(f). Black lines indicate the dispersion curve of the one-excitation manifold. Energy is in units of $\varepsilon$; wavenumber is in radians per lattice constant of the chromophore array. Calculations were performed on a $639 \times 480$ grid over $k \in \left[-10,10\right)$ $\times E \in \left[ 0,6 \right)$.}
\label{fig:13AbsDens}
\end{figure*}

\begin{figure*}[t]
\begin{center}
\includegraphics[scale=1.0]{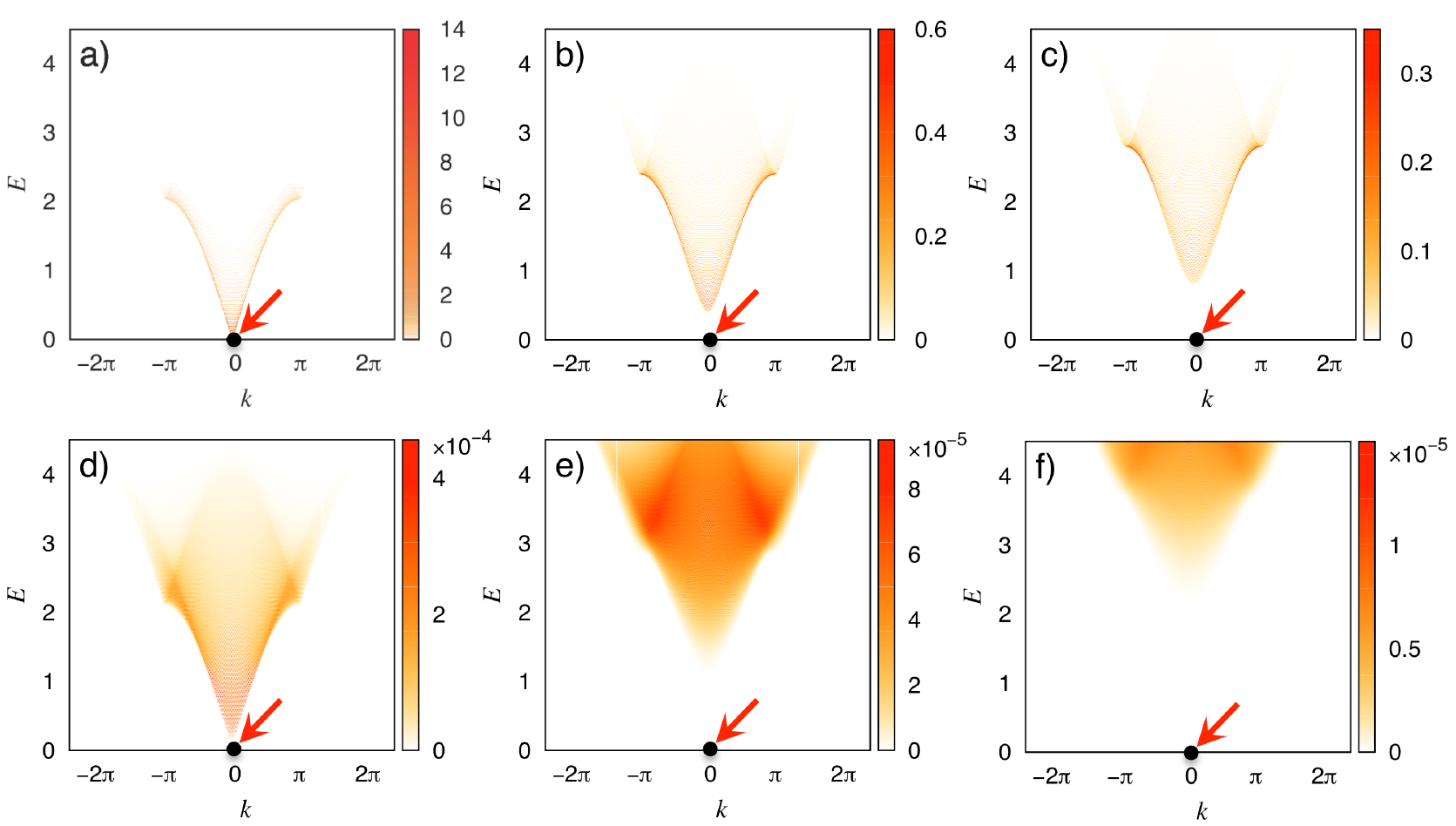}
\end{center}
\caption{Linear absorption density $\rho_A \left( k, E \right)$ from the ground state $\left| \Phi_+ \right\rangle$ of an array with $M=200$: to the two-excitation manifold, $G^\dagger \left( \alpha_1 \right) G^\dagger \left( \alpha_2 \right) \left|\Phi_{-} \right\rangle$ (a -- c) and to the four-excitation manifold, $G^\dagger \left( \alpha_1 \right) G^\dagger \left( \alpha_2 \right) G^\dagger \left( \alpha_3 \right) G^\dagger \left( \alpha_4 \right) \left| \Phi_{-} \right\rangle$ (d -- e), for $B =$ 1.02~(a,d), 1.2~(b,e), 1.4~(c,f). The black dot and red(gray) arrow indicate the $\left| \Phi_{+} \right\rangle \longrightarrow \left| \Phi_{-} \right\rangle$ transition. Energy is in units of $\varepsilon$; wavenumber is in radians per lattice constant of the chromophore array. Calculations were performed on a $639 \times 480$ grid over $k \in \left[-10,10\right)$ $\times E \in \left[ 0,6 \right)$.}
\label{fig:24AbsDens}
\end{figure*}

Absorption densities $\rho_A \left( k, E \right)$ for transitions from the KMM ground state, $\left| \Phi_{+} \right\rangle$, to higher excitation manifolds for values of $B=2 \left| b \right| / \varepsilon$ beyond those presented in Figures~\ref{fig:multiexc} and \ref{fig:fourexc} are shown in Figures~\ref{fig:13AbsDens} ($B < 1$) and \ref{fig:24AbsDens} ($B > 1$). The resolution is the same as for Figures~\ref{fig:multiexc} and \ref{fig:fourexc}. Note that for a given $B$ value the maximum absorption density in the four-excitation manifold is always several orders of magnitude smaller than the maximum absorption density in the two-excitation manifold.

\acknowledgments{A.A.K. acknowledges financial support of The Netherlands Organization for Scientific Research (NWO) Rubicon Grant 680-50-1022. K.B.W., J. D. and D.B.A. were supported by the DARPA QuBE program, B.A. and F. H. by NSF. We thank the Kavli Institute for Theoretical Physics for hospitality and for supporting this research in part by the National Science Foundation Grant No. PHY11-25915.}

\bibliography{Kocherzhenko_References.bib}

\begin{thebibliography}{39}
\bibitem[{not({\natexlab{a}})}]{note_othercases}
\bibinfo{note}{For $M$ even, all results for $b > 0$ may be obtained by
  applying a rotation about the $z$ axis through $\pi$ on either of the two
  sublattices. The case of $M$ odd is complicated by the periodic boundary
  conditions and will be considered elsewhere.}
\end{thebibliography}

\end{document}